\documentclass[9pt,twocolumn,twoside]{osajnl}
\usepackage{physics}
\journal{optica} % Choose journal (ao, aop, josaa, josab, ol, optica, pr)

\setboolean{shortarticle}{true}
\title{Fully-Quantum-Theoretic Numerical Study on \\
Quantum Phase Sensing and Ghost Imaging Systems Operating with Multimode N00N States}

\author[1]{Dong-Yeop Na}
\author[1]{Peter Bermel}
\author[1,*]{Weng Cho Chew}

\affil[1]{School of Electrical and Computer Engineering, Purdue University, West Lafayette IN, 47906}
\affil[*]{Corresponding author: wcchew@purdue.edu}

\begin{abstract}
We present a numerical study on the super-resolution of quantum phase sensing and ghost imaging systems operating with multimode N00N states beyond the Rayleigh diffraction limit. 
Our computational simulations are based on the canonical quantization via numerical mode-decomposition (CQ-NMD) \cite{Na2020quantum,Na2021Diagonalization}, in which normal (eigen) modes of electromagnetic fields in inhomogeneous dielectric media are numerically found using computational electromagnetics methods.
In the CQ-NMD framework and the Heisenberg picture, the expectation value of arbitrary observables with respect to initial quantum states of various non-classical lights can be evaluated with the use of Wick's theorem.
The present numerical framework has a great potential to deal with scattering problems of entangled photons due to arbitrary dielectric objects.
\end{abstract}

\setboolean{displaycopyright}{true}

\begin{document}

\maketitle

%%%%%%%%%%%%%%%%%%%%%%%%%%  body  %%%%%%%%%%%%%%%%%%%%%%%%%%
\section{Introduction}
The resolution of classical sensing and imaging systems is restricted by the Rayleigh diffraction limit that, roughly speaking, a subwavelength object less than half wavelength of light cannot be identified.
Quantum sensing and imaging technologies are of great interest since the use of entangled photons overcomes such fundamental resolution limit, viz., super-resolution.
Among various types of entanglement encoded in lights, the photon-number entanglement along different paths, called N00N states, is a promising candidate for quantum metrology.
Experimental works have verified the super-resolution in phase measurements using N00N states \cite{Dowling2008quantum} and shown the enhanced performance in quantum imaging systems \cite{Erkmen:10,Meda_2017}.
Furthermore, \cite{Manuel2018Super} has recently shown optical centroid measurement (OCM) of N00N states at the Heisenberg limit where the error in the phase measurement becomes $\Delta \theta \approx 1/N$.

In this letter, for the first time, we numerically demonstrate the super-resolution of quantum phase sensing and ghost imaging systems operating with multimode N00N states.
Our computational simulations employ the fully-quantum-theoretic numerical model, called the canonical quantization via numerical mode-decomposition (CQ-NMD) \cite{Na2020quantum,Na2021Diagonalization} based on the macroscopic quantum electrodynamics theory.
The CQ-NMD performs the canonical quantization of electromagnetic (EM) fields in inhomogeneous dielectric media \cite{Glauber1991quantum} via numerical normal (eigen) modes obtained using computational electromagnetic (CEM) methods.
In the CQ-NMD framework and the Heisenberg picture, one can evaluate the expectation value of arbitrary observables for initial quantum states of various non-classical lighs with the use of Wick's theorem \cite{PhysRev.80.268}.
Thus, the CQ-NMD approach is suited for quantum imaging, sensing, and radar applications, capable of dealing with scattering problems of entangled photons due to arbitrary dielectric objects.

%---------- Math/Physics model ----------%
\section{Fundamental Math/Physics Model}

\subsection{Quantum Maxwellian operator and quantum state}
Quantum optics describes the random behaviors of EM fields in the quantum regime.
To do this, one needs to perform the canonical quantization where classical Maxwellian field and source variables can be elevated into (infinite-dimensional) operators while quantum states are introduced spanning (infinite-dimensional) Hilbert spaces.
Quantum Maxwellian operators and quantum states are solutions to (1) quantum Maxwell's equations (QME) and (2) quantum state equation (QSE), respectively \cite{CHEW2016quantum,Chew2021quantum}.
This study adopts the Heisenberg picture where observable operators are time-dependent, whereas quantum states are not.
After performing the canonical quantization, one can express the positive frequency part of the quantized vector potential by the normal mode expansion, such as,
\begin{flalign}
\hat{\mathbf{A}}^{(+)}(\mathbf{r},t)
&=
\sum_{\lambda}
\int_{\Omega^{+}}
d\omega
\sqrt{\frac{\hbar}{2\omega}}
\boldsymbol{\Phi}_{\omega,\lambda}(\mathbf{r})
\hat{a}_{\omega,\lambda}e^{-i\omega t},
\label{eqn:vec_op_p}
\end{flalign}
where the hat  symbol $\hat{~}$  denotes an operator, $\hbar$ is the reduced Planck constant, $\omega$ is eigenfrequency, $\Omega^{+}$ is a set of positive eigenfrequencies, $\lambda$ denotes a generic degeneracy index, $\boldsymbol{\Phi}_{\omega,\lambda}(\mathbf{r})$ is time-harmonic vectorial normal mode for $(\omega,\lambda)$, and $\hat{a}_{\omega,\lambda}$ ($\hat{a}^{\dag}_{\omega,\lambda}$) is an annihilation (creation) operator.
The annihilation and creation operator obey the standard bosonic commutator relation.
Using the orthonormality of the normal modes, one can diagonalize the Hamiltonian operator written by $
\hat{H}=\sum_{\lambda}\int_{\Omega^{+}}d\omega \hbar\omega
\left(\hat{n}_{\omega,\lambda}
+
\frac{1}{2}\hat{I}
\right)
$ where $\hat{n}_{\omega,\lambda}=\hat{a}^{\dag}_{\omega,\lambda}\hat{a}_{\omega,\lambda}$ is called number operator.
The eigenstate of number operators are known as Fock states, i.e., $\hat{n}_{\omega,\lambda}\ket{n_{\omega,\lambda}}=n\ket{n_{\omega,\lambda}}$ where $\ket{n_{\omega,\lambda}}$ is the Fock state representing that $n$ number of photons are occupied in $(\omega,\lambda)$-th normal mode.
Thus, the mode-decomposition enables one to easily model arbitrary quantum states by the linear superposition of multimode Fock states.
One can refer to \cite{fox2006quantum} for more details.
\begin{figure}
\centering
\includegraphics[width=.8\linewidth]
{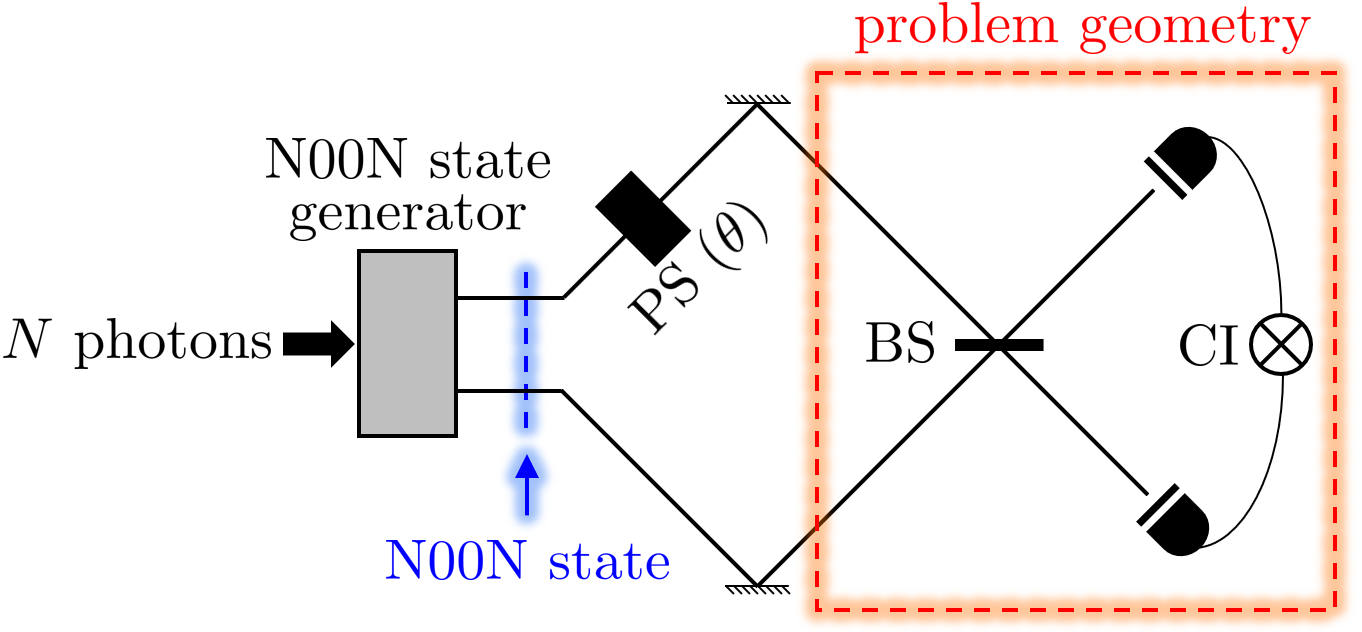}
\caption{Configuration of quantum phase sensing systems operating with N00N states. Note that BS, PS, CI represent beam splitter, phase shifter, coincidence counter, respectively.}
\label{fig:Fig_0}
\end{figure}

\subsection{Numerical mode-decomposition \cite{Na2020quantum,Na2021Diagonalization}}
To extract normal modes for EM fields in inhomogeneous dielectric media, one should solve the following vector wave equation
\begin{flalign}
\nabla\times \mu_{0}^{-1}\nabla\times \boldsymbol{\Phi}_{\omega,\lambda}(\mathbf{r})
-
\omega^{2}\epsilon(\mathbf{r})
\boldsymbol{\Phi}_{\omega,\lambda}(\mathbf{r})
=0.
\label{eqn:Helmholtz_WE}
\end{flalign}
Note that this study considers lossless and dispersionless dielectric media for simplicity.
To model arbitrary geometric and medium complexity in dielectric scatterers, we can utilize numerical methods in CEM such as finite-difference or finite-element methods.
The resulting discrete counterpart of \eqref{eqn:Helmholtz_WE} with Bloch periodic boundary conditions becomes a finite-dimensional generalized Hermitian eigenvalue problem written by $
\overline{\mathbf{S}}
\cdot
\overline{\boldsymbol{\Phi}}
=
\overline{\mathbf{M}}
\cdot
\overline{\boldsymbol{\Phi}}
\cdot \overline{\boldsymbol{\omega}}^{2}
$ where $\overline{\mathbf{S}}$ and $\overline{\mathbf{M}}$ are (sparse) stiffness and mass matrices, which encodes double-curl operator and medium and metric information, $\overline{\boldsymbol{\Phi}}$ is a (full) matrix including numerical normal modes, and $\overline{\boldsymbol{\omega}}$ is a diagonal matrix including relevant eigenfrequencies.

\subsection{Representation based on CQ-NMD framework}
With the use of numerical normal modes, the continuum modal index $(\omega,\lambda)$ is replaced by a single discrete modal index $i$.
Thus, the resulting vector potential operator at $j$-grid point, denoted by $\mathbf{r}_{j}$, can be rewritten by
\begin{flalign}
\hat{\mathbf{A}}^{(+)}(\mathbf{r}_{j},t)
\approx
\sum_{i=1}^{N_{\text{dof}}}
\sqrt{\frac{\hbar}{2\omega_i}}
\boldsymbol{\Phi}_{i}(\mathbf{r}_{j})
\hat{a}_{i}e^{-i \omega_{i} t}
\label{eqn:numerical_field_operator}
\end{flalign}
where $N_{\text{dof}}$ denotes the total number of numerical normal modes.
The resulting Hamiltonian can be also written by $
\hat{H}\approx\sum_{i=1}^{N_{\text{dof}}}
\hbar\omega_{i}
\left(\hat{a}_{i}^{\dag}\hat{a}_{i}+\frac{1}{2}\hat{I}\right).$
\begin{figure}
\centering
\includegraphics[width=.8\linewidth]
{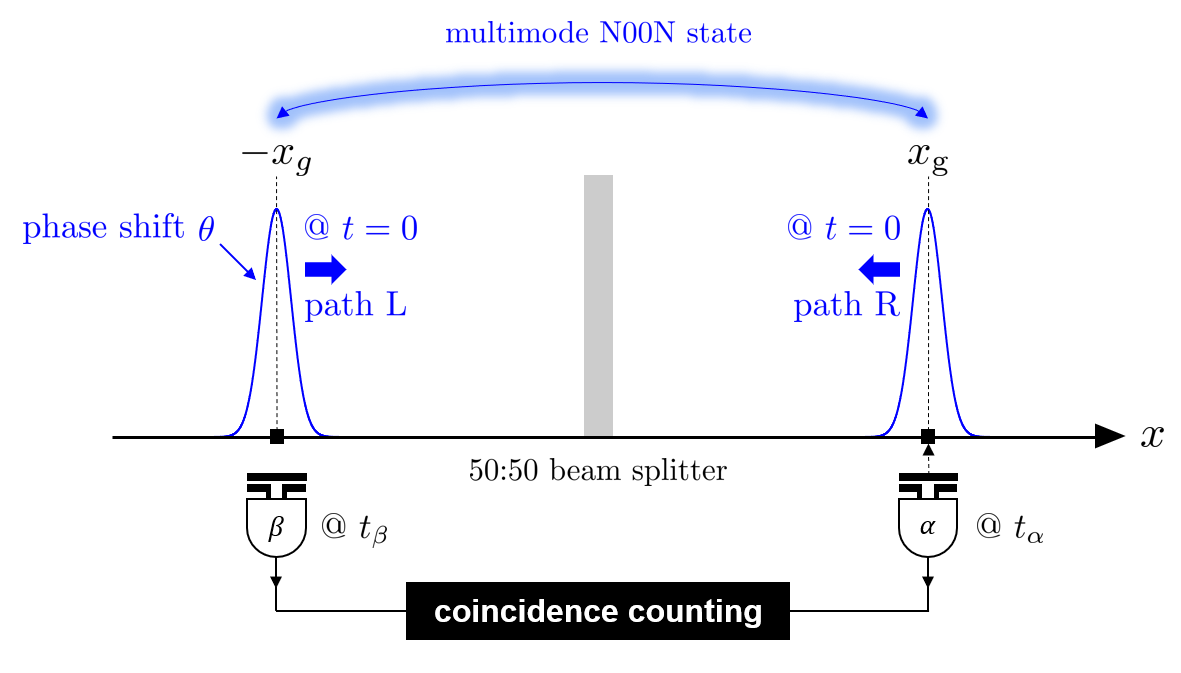}
\caption{The problem geometry of one-dimensional simulations of the phase sensing system described in Fig. \ref{fig:Fig_0} to observe the enhanced phase measurement sensitivity.}
\label{fig:Fig_1}
\end{figure}

\subsection{Modeling multimode N00N states}
A typical N00N state takes the form of
\begin{flalign}
\ket{\psi}_{\text{N00N}}
=
\frac{1}{\sqrt{2}}
\Bigl(
\ket{\psi_{1}^{(N)}}\ket{\emptyset_{2}}
+
\ket{\emptyset_{1}}\ket{\psi_{2}^{(N)}}
\Bigr)
\end{flalign}
where $\ket{\emptyset_{m}}$ and $\ket{{\psi_{m}^{(N)}}}$ represent quantum states of no-photon and $N$ number of (monochromatic) photons along $m$-th path for $m=1,2$, respectively.
Furthermore, we assume that each photon is riding on a wavepacket (i.e., quasi-monochromatic) such that the resulting quantum state should be expanded by multimode Fock states \cite{Mandel1995optical} $
\ket{\psi^{(1)}_{m}}
\approx
\sum_{i=1}^{N_{\text{dof}}}
\tilde{g}^{(m)}_{i}
\hat{a}^{\dag}_{i}\ket{0}
$ where $\tilde{g}_{i}^{(m)}$ represents a probability amplitude of $i$-th single-photon Fock state that encodes the spectrum of a wavepacket along $m$-th path.
Similarly, a quantum state of $N$ quasi-monochromatic photons occupied in $m$-th path can be explicitly expressed by
\begin{flalign}
\ket{\psi_{m}^{(N)}}
\approx
\frac{1}{\sqrt{N}!}
\left(
\sum_{i=1}^{N_{\text{dof}}}
\tilde{g}^{(m)}_{i}
\hat{a}^{\dag}_{i}
\right)^{N}
\ket{0}.
\label{eqn:poly_N_photon_state}
\end{flalign}
Note that the multimode N00N state fulfills the normalization condition rigorously, proven by using Wick's theorem (See the supplementary material).

\subsection{Modeling coincidence counting}
Photon statistics is the theoretical and experimental study to identify the statistical distributions of photons produced in a light source via photon counting experiments.
Particularly, coincidence counting, referring to the simultaneous detection of two or more photons at photodetectors, is of cardinal importance in quantum optics widely used to study the quantum state of non-classical lights.
Here, we define N-th order correlation function (CF) for a pair of photodetectors (indexed by $\alpha$ and $\beta$) \cite{Mandel1995optical} as 
\begin{flalign}
\text{N-th order CF}=
\frac{
\mel
{\psi}
{
\hat{\alpha}^{(-)}_{l}
\hat{\beta}^{(-)}_{l}
\hat{\beta}^{(+)}_{l}
\hat{\alpha}^{(+)}_{l}
}
{\psi} 
}{
\mel
{\psi}
{
\hat{\alpha}^{(-)}_{l}
\hat{\alpha}^{(+)}_{l}
}
{\psi} 
\mel
{\psi}
{
\hat{\beta}^{(-)}_{l}
\hat{\beta}^{(+)}_{l}
}
{\psi} 
}
\label{eqn:coincidence_1}
\end{flalign}
where $\ket{\psi}$ is an initial quantum state, and, for $\zeta=\alpha$ or $\beta$, $\hat{\zeta}^{(-)}_{l}=\left(\hat{\zeta}^{(+)}_{l}\right)^{\dag}$ and $
\hat{\zeta}^{(+)}_{l}
=
\prod_{i=1}^{N/2}
\left[\hat{\mathbf{A}}^{(+)}(\mathbf{r}_\zeta,t_\zeta)\right]_{l}
$ where $\mathbf{r}_{\zeta}$ and $t_{\zeta}$ denote $\zeta$-th photodetector's location and time, respectively, and subscript $l=x,y,z$ denotes a component of a vectorial field operator $\hat{\mathbf{A}}^{(+)}$.
Physically speaking, the numerator represents N-fold coincidence count and two terms in the denominator are normalization factors, which are associated with the photodetection probability at each photodetector independently.
Note that for simplicity we assume the photodetection number at each photodetector to be equal, i.e., N/2-photodetection per photodetector.\footnote{This is one of possible configurations to calculate the N-order CF. Unlike the case $N=2$, if $N$ is large, there are many possible photodetection configuations for the N-order CF when using a pair of photodetectors. For example, when $N=4$ for binary paths (say $\alpha$ and $\beta$), possible non-entangled quantum states are $\ket{1}_{\alpha}\ket{3}_{\beta}$, $\ket{2}_{\alpha}\ket{2}_{\beta}$, and $\ket{3}_{\alpha}\ket{1}_{\beta}$; therefore, one needs to perform (1,3)-, (2,2)-, (3,1)-times photodetections at photodetectors $\alpha$ and $\beta$ to identify above non-entangled quantum states, respectively.}
To calculate \eqref{eqn:coincidence_1}, we translate it to products of ladder operators by substituting $\ket{\psi}_{\text{N00N}}$ and \eqref{eqn:numerical_field_operator} into \eqref{eqn:coincidence_1}.
Then, one can apply Wick's theorem \cite{PhysRev.80.268} to have the normal order of the products of ladder operators and then sum up full-contraction terms (See the supplementary material for the details).
\begin{figure}
\centering
\includegraphics[width=0.8\linewidth]
{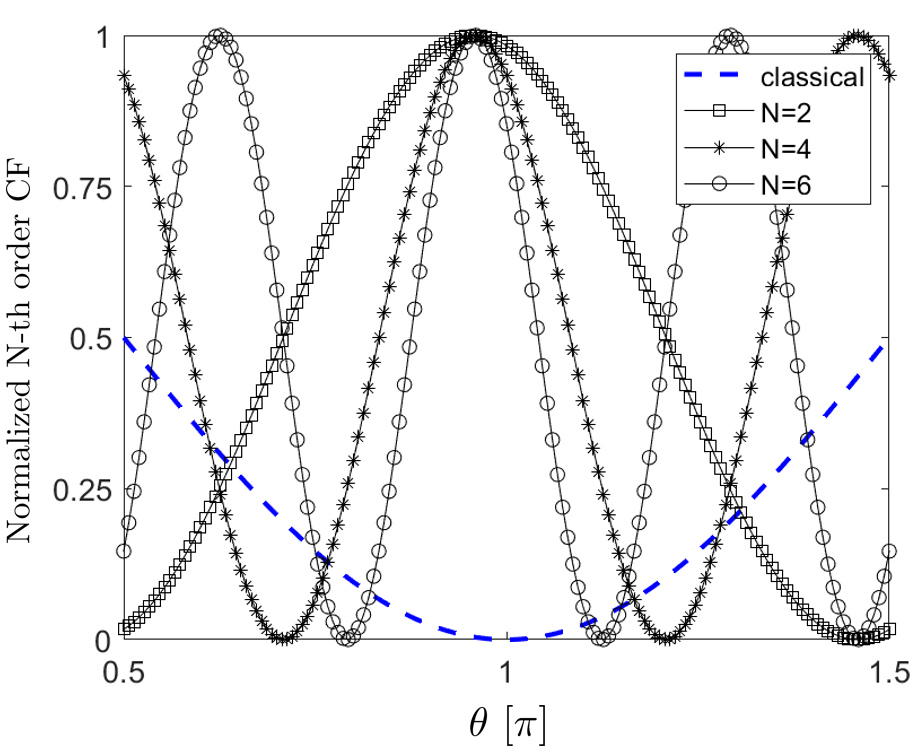}
\caption{N-th order correlation function (CF) versus $\theta$ (phase shifter) for $N=2,4,6$, compared with the classical case.}
\label{fig:Fig_2}
\end{figure}

\section{Simulation results}
\subsection{Quantum phase sensing system}\label{subsec_phase_sim}
Consider a quantum phase sensing system, consisting of a N00N state generator, phase shifter, beam splitter, and coincidence counting measurement circuit, as illustrated in Fig. \ref{fig:Fig_0}.
The phase shifter ($\theta$), inserted on the upper path, adds the phase $\theta$ into the probability amplitude of a quantum state on that arm, and the divided N00N state is self-interfered through a 50:50 beam splitter.
We can observe a correlation pattern with respect to $\theta$ from calculating \eqref{eqn:coincidence_1}.
Here, instead of modeling the entire quantum phase sensing system, we consider the beam splitter part only with a proper initial N00N state that incorporates the phase shifter effect.
Note that we assume that the overall interaction time in the phase sensing system is much below the dephasing time T2 of N00N states (see the supplementary document including the effect of T2 to the super-resolution).
The one-dimensional problem geometry is illustrated in Fig. \ref{fig:Fig_1}.
A multimode N00N state including N number of quasi-monochromatic photons is initialized on the left and right, taking the form of
\begin{flalign}
\ket{\psi}_{\text{IN}}
&=
\frac{1}{\sqrt{2 N!}}
\left[
\left(
e^{i\theta}
\sum_{i=1}^{N_{\text{dof}}}
\tilde{g}^{(L)}_{i}
\hat{a}^{\dag}_{i}
\right)^{N}\ket{0}
+
\left(
\sum_{i=1}^{N_{\text{dof}}}
\tilde{g}^{(R)}_{i}
\hat{a}^{\dag}_{i}
\right)^{N}\ket{0}
\right].
\nonumber
\end{flalign}
Note that the addition of $e^{i\theta}$ factor in the first term takes into account the phase shifter effect. 
Two incident wavepackets, initially localized at $\pm x_{g}$ where $x_g = 0.375$ [m], respectively, are supposed to travel toward the 50:50 beam splitter in the center.
The shape of wavepackets was assumed to be Gaussian.
The center frequency and deviation of the Gaussian wavepacket are $\omega_{g}=526c$ and $\sigma_g=1.59c$, respectively.
After the interference of photons in the beam splitter, we calculated N-th order CF in \eqref{eqn:coincidence_1}.
Fig. \ref{fig:Fig_2} shows normalized N-th order CF versus $\theta$ for N=2,4,6.
It is observed that normalized N-th order CF oscillates N-times faster than that of the coherent state (classical limit), i.e., the peak spacing when using multimode N00N states becomes $\lambda_{\text{N00N}}=\lambda_{\text{coherent}}/N$ where $\lambda_{\text{coherent}}$ is of the coherent state, exhibiting the super-resolution. 
The simulation results agree with the theoretical prediction \cite{Dowling2008quantum}.
This super-resolution comes from the fact that when a pure photon number state $\ket{N}$ passes through a phase shifter $\theta$, the coherent accumulation of the phase shift $\theta$ experienced by each single photon is possible.
The net phase delay, i.e., $e^{i N\theta}$, is then transformed into the probability amplitude of a N00N state.
Thus, correlation patterns in the Mach-Zehnder interferometer can oscillate depending the net phase delay.
On the other hand, the action of the phase shifter on a coherent (classical) state with the average N photons averages out all phase shifts experienced by different photon number states (due to the linear superposition); consequently, the probability amplitude can only gain the phase delay $e^{i\theta}$.
One can find more details in supplementary material.

\subsection{Quantum Ghost-Imaging}
\begin{figure}
\centering
\includegraphics[width=.8\linewidth]
{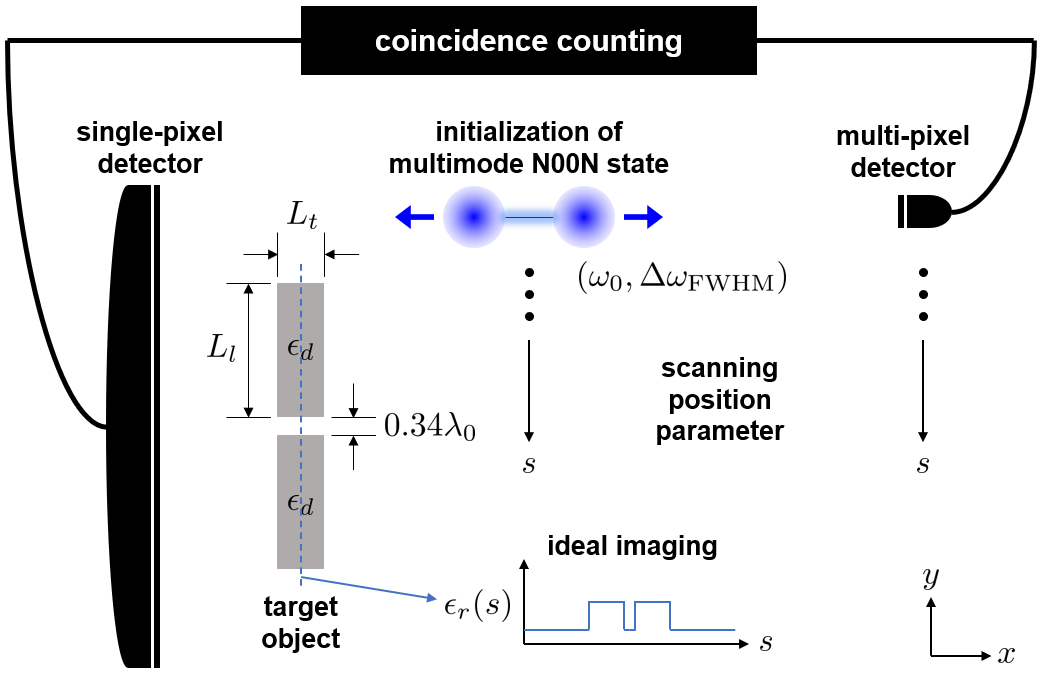}
\caption{Two-dimensional simulation scenario of a quantum ghost imaging system using a multimode N00N state.}
\label{fig:Fig_3}
\end{figure}

\begin{figure}
\centering
\includegraphics[width=.8\linewidth]
{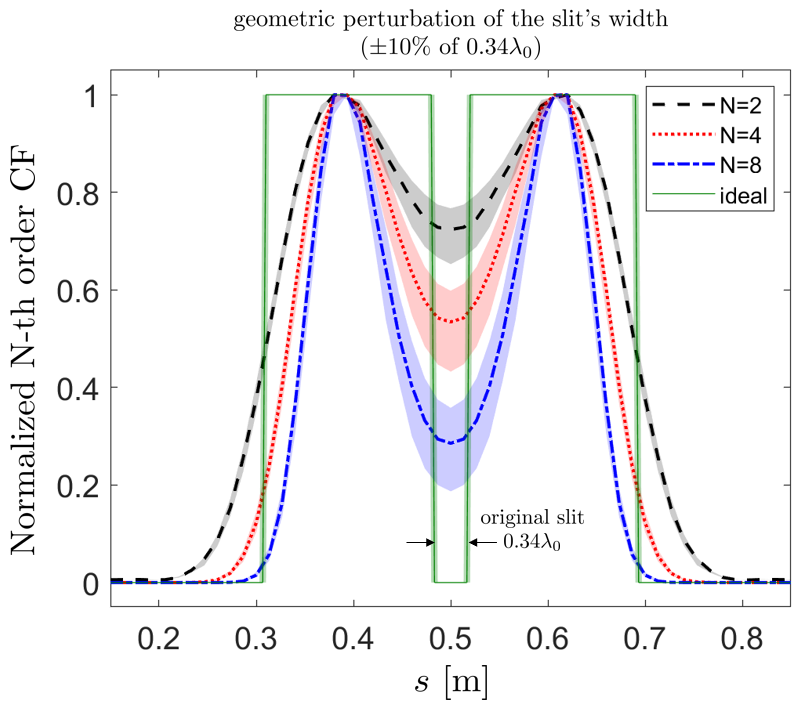}
\caption{N-th order correlation function (CF) versus scanning parameter $s$ for $N=2,4,8$, compared with the reference.}
\label{fig:Fig_4}
\end{figure}

Next, consider a quantum ghost imaging system whose two-dimensional simulation scenario is illustrated in Fig. \ref{fig:Fig_3}.
A multimode N00N state including $N$ quasi-monochromatic photons is initialized in the middle.
The center frequency and full width at half maximum (FWHM) of each photon are assumed to be $\omega_0=50c$ [rad/s] ($\lambda_0\approx 1.26\times 10^{-1}$ [m]) and $\Delta \omega_{\text{FWHM}}=1.56c$ [rad/s].
We assume that vector potential operators are polarized along $z$ direction.
An object to be imaged is a dielectric slab (the relative dielectric constant is $\varepsilon_d=4$.) including a subwavelength slit.
The slit width $L_g=0.34\lambda_0=4.33\times 10^{-2}$ [m], the length of each side of the dielectric slab $L_l=1.70\times 10^{-1}$ [m], and the slab thickness $L_t=9.67\times 10^{-2}$ [m].
We place a single-pixel (or bucket) photodetector on the left behind the object while locating a multi-pixel photodetector on the right.
The entangled photons propagating to the left hit the target object and are collected at the single-pixel detector. In contrast, the rest entangled-photon propagating toward the right is measured at the multi-pixel detector without having any interference with the target object.
It is assumed that each photodetector has the photon number resolving capability.
Repeating the above tasks at each scanning position parameter $s$ (see Fig. \ref{fig:Fig_4}), we can calculate the N-th order CF in terms of $s$ for various $N$.
Fig. \ref{fig:Fig_4} shows normalized N-th order CF versus scanning position parameter $s$ for $N=2,4,8$.
In an ideal case, as illustrated by a green-solid-line in Fig. \ref{fig:Fig_4}, N-th order CF should be zero in the absence of the object; otherwise, it is to be unity.
This behavior can be deduced from the definition of N-th order CF in \eqref{eqn:coincidence_1}.
The numerator measures the degree of coincidence, whereas terms in the denominator measure the photodetection probability at each photodetector independently.
In the absence of the dielectric object, original entanglement is preserved, implying that the numerator goes to zero due to no coincidence count but terms in the denominator can have non-zero values (50\% chance of photodetection for each).
On the other hand, when a very high contrast (hard) object is present, photons may not pass through them.
This also makes the coincidence count zero, however, one of terms in the denominator becomes zero as well; consequently, the N-th order CF become unity in the limit of zero divided zero \cite{Na2020Classical}.
When an object is soft and arbitrary shaped, this beam-physics-based argument may not work simply but photon fields experience a more complicated mode conversion process.
Or, when components are integrated into a much smaller volume, the full-wave physics plays an important role.
In these cases, the mode conversion process should be taken into account to quantify the N-th order CF more accurately.
Also, the mode conversion process is imperative when considering higher-dimensional cases (two- or three-dimensional) since higher dimensional spaces have a much larger degeneracy space, in particular, studying scattering of (non-local) quantum states of light in an ambient space, such as quantum radar and imaging systems, rather than well-confined optical fibers.
Note that such mode conversion process is embedded in numerical normal modes implicitly.  
The simulation results show that increasing the photon number results in the $N$-times higher resolution.
We also studied the effect of the geometric perturbation in the slit width on the imaging result.
Specifically, we perturb the slit width by $\pm 10\%$ of the original one.
The resultant imaging results are illustrated by shaded regions based on the unperturbed results (solid lines).
It is interesting to observe that the geometric perturbation significantly affects the imaging results around the slit region, especially when the photon number is higher.
This implies that the performance of quantum ghost imaging systems operating with high N00N states may be highly sensitive to small perturbations when imaging subwavelength objects.

\section{Conclusion}
We have performed fully-quantum-theoretic computational simulations of quantum phase sensing and ghost imaging systems operating with multimode N00N states to observe the super-resolution based on the canonical quantization via numerical mode-decomposition (CQ-NMD) approach \cite{Na2020quantum,Na2021Diagonalization}.
The simulation results agreed well with both theoretical estimates and experimental observations that the increase of the photon number $N$ achieves $N$-times higher sensitivity and resolution.
The present study has shown great promise of utilizing the conventional computational electromagnetic methods together with quantum Maxwell's equations and quantum state equation for quantum metrology applications.
Although our simulations assumed an ideal condition, three practical issues need to be resolved for fully taking the quantum advantages: (1) generating arbitrary high N00N states, (2) having the photon-number-resolving photodetection capability and (2) improving the extreme fragility to the interaction with environment \cite{PhysRevA.80.013825}.
For our future studies, we plan to account for dissipation and dispersion effects of media on the performance of quantum sensing and imaging systems.
Moreover, we plan to consider various kinds of non-classical states of light for quantum metrology applications, such as, squeezed states, entangled coherent states \cite{PhysRevLett.107.083601}, and N00N-like states \cite{PhysRevA.95.032321}, which are relatively easier to create as well as detect in practice while still exhibiting the quantum advantages from the metrology aspect.

% Bibliography
\bibliography{sample}

% Full bibliography added automatically for Optics Letters submissions; the following line will simply be ignored if submitting to other journals.
% Note that this extra page will not count against page length
\bibliographyfullrefs{sample}

%Manual citation list
%\begin{thebibliography}{1}
%\bibitem{Zhang:14}
%Y.~Zhang, S.~Qiao, L.~Sun, Q.~W. Shi, W.~Huang, %L.~Li, and Z.~Yang,
 % \enquote{Photoinduced active terahertz metamaterials with nanostructured
  %vanadium dioxide film deposited by sol-gel method,} Opt. Express \textbf{22},
  %11070--11078 (2014).
%\end{thebibliography}

% Please include bios and photos of all authors for aop articles
\ifthenelse{\equal{\journalref}{aop}}{%
\section*{Author Biographies}
\begingroup
\setlength\intextsep{0pt}
\begin{minipage}[t][6.3cm][t]{1.0\textwidth} % Adjust height [6.3cm] as required for separation of bio photos.
  \begin{wrapfigure}{L}{0.25\textwidth}
    \includegraphics[width=0.25\textwidth]{john_smith.eps}
  \end{wrapfigure}
  \noindent
  {\bfseries John Smith} received his BSc (Mathematics) in 2000 from The University of Maryland. His research interests include lasers and optics.
\end{minipage}
\begin{minipage}{1.0\textwidth}
  \begin{wrapfigure}{L}{0.25\textwidth}
    \includegraphics[width=0.25\textwidth]{alice_smith.eps}
  \end{wrapfigure}
  \noindent
  {\bfseries Alice Smith} also received her BSc (Mathematics) in 2000 from The University of Maryland. Her research interests also include lasers and optics.
\end{minipage}
\endgroup
}{}

\end{document}